\documentclass[12pt,epsfig]{article}          
\usepackage{epsfig}
\makeatletter  

\@addtoreset{equation}{section} \makeatother

\begin{document}

\vspace*{0.5cm}

\newcommand{\hs}{\hspace*{0.5cm}}
\newcommand{\vs}{\vspace*{0.5cm}}
\newcommand{\be}{\begin{equation}}
\newcommand{\ee}{\end{equation}}
\newcommand{\bea}{\begin{eqnarray}}
\newcommand{\eea}{\end{eqnarray}}
\newcommand{\nn}{\nonumber}
\newcommand{\crn}{\nonumber \\}
\newcommand{\non}{\nonumber}
\newcommand{\noi}{\noindent}
\newcommand{\al}{\alpha}
\newcommand{\la}{\lambda}
\newcommand{\bet}{\beta}
\newcommand{\ga}{\gamma}
\newcommand{\va}{\varphi}
\newcommand{\om}{\omega}
\newcommand{\pa}{\partial}
\newcommand{\fr}{\frac}
\newcommand{\bc}{\begin{center}}
\newcommand{\ec}{\end{center}}
\newcommand{\Ga}{\Gamma}
\newcommand{\de}{\delta}
\newcommand{\De}{\Delta}
\newcommand{\ep}{\epsilon}
\newcommand{\varep}{\varepsilon}
\newcommand{\ka}{\kappa}
\newcommand{\La}{\Lambda}
\newcommand{\si}{\sigma}
\newcommand{\Si}{\Sigma}
\newcommand{\ta}{\tau}
\newcommand{\up}{\upsilon}
\newcommand{\Up}{\Upsilon}
\newcommand{\ze}{\zeta}
\newcommand{\ps}{\psi}
\newcommand{\Ps}{\Psi}
\newcommand{\ph}{\phi}
\newcommand{\vph}{\varphi}
\newcommand{\Ph}{\Phi}
\newcommand{\Om}{\Omega}

\newcommand {\ba}{\begin{array}}
\newcommand {\ea}{\end{array}}

\bc {\Large Scalar sector of   Supersymmetric\\

 $\mbox{SU}(3)_C\otimes \mbox{SU}(3)_L \otimes \mbox{U}(1)_N$
 Model\\

  with right-handed neutrinos}\\
\vspace*{1cm}

{\bf D. T. Huong$^a$, M. C. Rodriguez$^b$} and {\bf H. N. Long$^a$}\\

\vspace*{0.5cm}

$^a$ {\it Institute of Physics, VAST, P. O. Box 429, Bo Ho, Hanoi
10000, Vietnam}\\

 $^b$ {\it Funda\c{c}\~{a}o Universidade Federal do Rio Grande-FURG,
 Departamento de F\'\i sica,\\
 Av. It\'alia, km 8, Campus Carreiros,
 96201-900 Rio Grande, RS,  Brazil}

\ec

\begin{abstract}
We investigate a scalar sector of the supersymmetric $ SU_C(3)
\otimes SU_L(3)\otimes U_N(1)$ model with right-handed neutrinos.
The mass spectra are derived. We show that only neutral Higgs
sector with lepton number $L=0$ could have a VEV. There is no
mixing between scalars having $L=0$ and  bilepton scalars having
$L=2$. There are six Goldstone bosons:  two in neutral sector,
three in pseudo-scalar sector and one in
 charged scalar sector. For a given set of input parameters (five from
the  $F$ terms and two from the soft term)
 all the scalar sectors in this model
contain the upper limit of 230 GeV to the mass of the lightest
scalar, which are in agreement with the lower limit of the SM
Higgs boson obtained by LEP.

\end{abstract}

PACS number(s): 12.60. Jv, 12.60.Fr.


\section{Introduction}

The models based on the $ \mbox{SU}(3)_C\otimes \mbox{SU}(3)_L
\otimes \mbox{U}(1)_X$ (called 3-3-1 models)
\cite{singer,ppf,331rh} provide possible solutions to some puzzles
of the standard model (SM) such as the generation number problem,
the electric charge quantization \cite{dongl2}. Since one
generation of quarks is treated differently from the others this
may  lead to a natural explanation for  the large mass of the top
quark~\cite{longvan}. These models also furnish a good candidate
for self-interacting dark matter (SIDM) since there are two Higgs
bosons, one scalar and one pseudoscalar, which have the properties
of candidates for dark matter like stability, neutrality and that
it must not overpopulate the universe~\cite{longlan}, etc.

There are two main versions of the 3-3-1 models as far as lepton
sector is concern. In the minimal version, the charge conjugation
of the right-handed charged lepton for each generation is combined
with the usual $SU(2)_L$ doublet left-handed leptons components to
form an $SU(3)$ triplet $(\nu, l, l^c)_L$.  No extra leptons are
needed and there we shall call such models minimal 3-3-1 models.
There is no right-handed (RH) neutrino in its minimal version.
Another version adds a left-handed anti-neutrino to each usual
$SU(2)_L$ doublet left-handed lepton to form a triplet. This model
is called the 3-3-1 model with RH  neutrinos.

The supersymmetric version of the model of Ref.\cite{331rh} has
already  been presented in Ref.\cite{s331r}.  However, the authors
of  Ref.\cite{s331r} have just mentioned on the neutral Higgs sector. The
Higgs sector still remains one of the most indefinite part of the
SM, but it still represents a fundamental rule by explaining how
the particles gain masses by means of an isodoublets scalar field,
which is responsible for the spontaneous breakdown of the gauge
symmetry, the process by which the spectra of all particles are
generated.The Higgs mechanism plays a central role in gauge
theories. In this paper we will consider in detail the Higgs
sector of the model.

This paper is organized as follows. In Sec. \ref{sec1} we review
the non-supersymmetric 3-3-1 model with RH neutrinos and introduce
the respective superpartners. The construction of the
supersymmetric scalar potential is discussed in Sec.
\ref{sec:mass}, while in Sec. \ref{sec:massspectrum} we derive the
mass spectrum of the scalar sector of the model. The numerical
analysis are given in Sec. \ref{sec:numerical} and our plots are
presented in Sec. \ref{sec:plots}. Finally, the last section is
devoted to our conclusions.

\section{A review of the model}
\label{sec1}

In this section we present one review of the model we will consider on this
work.

\subsection{The non-supersymmetric 331 model with RH neutrinos}
\label{sec1a}

Let us first summarize the non-supersymmetric model \cite{331rh}. The leptons
transforming under the 3-3-1 factors as
\begin{equation}
L_{aL} =
\left( \begin{array}{c} \nu_a \\
                        l_a \\
                        \nu^{c}_{a}
\end{array} \right)_{L} \sim ( {\bf 1},{\bf 3},-1/3),
 \label{trip}
\end{equation}
with $a= e, \mu, \tau $ and $\nu_{a}^{c}=C \bar{ \nu_{a}}^{T}$,
plus the singlets
\begin{eqnarray}
l^{c}_{aL} \sim ({\bf 1},{\bf 1},1). \label{singletos}
\end{eqnarray}

In the quark sector we have the first two families transforming as
antitriplets of $SU(3)_L$
\begin{equation}
Q_{\alpha L} =
\left( \begin{array}{c} d_\al \\
                        u_\al \\
                        D_\al \end{array} \right)_{L} \sim
\left( {\bf 3},{\bf 3^{*}},0 \right), \;\;\alpha=1,2;
\label{quark323}
\end{equation}
with the respective singlets
\begin{equation}
u^{c}_{\alpha L} \sim \left( {\bf 3^{*}},{\bf 1},-2/3 \right),
\;\; d^{c}_{\alpha L},\; D^c_{\alpha L}  \sim \left( {\bf
3^{*}},{\bf 1},1/3 \right). \label{sing1}
\end{equation}

The third family transforms as triplet under $SU(3)_L$
\begin{eqnarray}
Q_{3L} &=&
\left( \begin{array}{c} u_3 \\
                        d_3 \\
                        T
\end{array} \right)_{L} \sim ( {\bf 3},{\bf 3},1/3),
\label{quarks3}
\end{eqnarray}
and their respective singlets
\begin{equation}
u^{c}_{3L},\; T^{ c}_{L} \sim ( {\bf 3^{*}},{\bf 1},-2/3), \;\;
d^{c}_{3L} \sim ( {\bf 3^{*}},{\bf 1},1/3 ). \label{qsingletos}
\end{equation}

In the scalar sector only two triplets $\eta\sim({\bf1}, {\bf3},
-1/3)$ and $\rho\sim({\bf1},{\bf3}, 2/3)$ are necessary to break
appropriately the gauge symmetry and also to give the correct mass
to all the fermions in the model. However, to eliminate flavor
changing neutral currents we add an extra scalar triplet
transforming like $\eta$. Therefore, the scalars of our model are written
as
\begin{eqnarray}
\eta &=&
\left( \begin{array}{c} \eta^{0}_{1} \\
                        \eta^{-} \\
                        \eta^{0}_{2}
\end{array} \right), \;\;
\chi =
\left( \begin{array}{c}  \chi^{0}_{1} \\
                        \chi^{-} \\
                        \chi^{0}_{2}
\end{array} \right) \sim ( {\bf 1},{\bf 3},-1/3 ),
\nonumber \\
\rho &=&
\left( \begin{array}{c}  \rho^{+}_{1} \\
                        \rho^{0} \\
                        \rho^{+}_{2}
\end{array} \right) \sim ( {\bf 1},{\bf 3},2/3 ),
\label{esctrip}
\end{eqnarray}
and we will denote the vacuum expectation values which are
different from zero as $v=\langle\eta^0_1\rangle$,
$w=\langle\chi^0_2\rangle$ and $u=\langle\rho^0\rangle$.

Despite $\eta$ and $\chi$ have the same quantum number, but they
members are quite different \cite{dias}: \bea L (\eta^{0}_{1},
\eta^{-},  \rho^{+}_{1}, \rho^{0},  \chi^{0}_{2}) & = & 0,\crn L(
\eta^{0}_{2},  \rho^{+}_{2},  \chi^{0}_{1}, \chi^{-}) &=& 2
\label{leptonnscalar} \eea

 The $ \eta^0_{2}$ and $\chi^{0}_{1}$ are scalar bileptons
$L(\eta^0_{2},\chi^{0}_{1})= 2$, while
$\rho^0, \chi^{0}_{2}$ do not have lepton number $L(\rho^0,
\chi^{0}_{2})= 0$~\cite{dias}. It is to be noted that, only pure
({\it without lepton number}) neutral scalars can have VEVs.

With this mention, one expect that  the would-be Goldstone bosons
coming from the Higgs potential can be written as:
 \be
\rho = \left( \ba {c} G_W^+ \\
u + i G_Z \\ 0 \ea \right), \hs
 \eta = \left( \ba{c} v + i G_Z
  \\  -G_W^-\\ 0 \ea \right), \hs
\chi = \left( \ba{c}G^{0}_X  \\ G^-_Y \\
 w+i G_{Z'} \ea \right)
\ee where $G_W^{\pm}$, $G_Z$, $G_Y^\pm$ and $G_{X}^{0}$, $G_{X}^{0
*}$ are the would-be Goldstone bosons  for the fields $W^\pm$,
$Z$, $Y^\pm$ and $X^{0}$,  $X^{0 *}$, respectively. The $\rho$ and
$\eta$ components give origin to  the charged Higgs $H^+$,
odd-$A^0$, even-$H^0$ (all with  masses of the order of $w$, scale
of energy of the first symmetry breaking) and the light Higgs
$h^0$ coming from the electroweak scale.
 The Higgs fields $\rho^{+}_2$, $\eta_2^0$ and $\chi^0_2$ have a mass
proportional to the scale $w$.  The other fields of $\chi$ give
origin to the would-be Goldstone bosons of $X^{0}$, $Y^-$ and
$Z'$.  All the scalar fields, except for $h^0$, have masses of the
order of the first symmetry breaking $M_\chi$.

\subsection{Supersymmetric partners}
\label{subsec:susyp}

Now, we introduce the minimal set of particles in order to implement the
supersymmetry. Here we will follow the usual notation writing for a given
fermion $f$, the respective sfermions by $\tilde{f}$ {\it i.e.},
$\tilde{l}$ and $\tilde{q}$ denote sleptons and squarks
respectively~\cite{haber}. Then, we have the following additional
particles

\begin{eqnarray}
\tilde{Q}_{\alpha L}& =&
\left( \begin{array}{c} \tilde{d}_\alpha \\
                        \tilde{u}_\alpha \\
                        \tilde{D}_\alpha
         \end{array}\right)_L
\sim ( {\bf 3},{\bf 3^{*}},0 ), \; \tilde{Q}_{3L} =
\left( \begin{array}{c} \tilde{u}_3 \\
                        \tilde{d}_3 \\
                        \tilde{T}
         \end{array}
\right)_{L}\!\!\!\! \sim ( {\bf 3},{\bf 3},1/3 ), \nonumber \\
\tilde{L}_{aL} &=&
\left( \begin{array}{c} \tilde{ \nu}_{a} \\
                        \tilde{l_a} \\
                        \tilde{ \nu}^{c}_{a}
\end{array} \right)_{L} \sim ( {\bf 1},{\bf 3},-1/3 ),
\label{susytri}
\end{eqnarray}

\begin{eqnarray}
\lefteqn{\tilde{l}^{c}_{aL}\sim ({\bf 1},{\bf 1},1),} \nonumber \\
&& \tilde{u}^{c}_{i L},\,\tilde{T}^{ c}_{L} \sim ( {\bf 3}^*,{\bf
1},-2/3),\;
\tilde{d}^{c}_{iL}, \tilde{D}^{  c}_{\alpha L}
 \sim ( {\bf 3}^*,{\bf 1},1/3),
\label{ss}
\end{eqnarray}
with $a=e, \mu , \tau$; $i=1,2,3$; and $\alpha=1,2$. However, when
considering quark (or squark) singlets of a given charge we will
use the notation $u^c_{iL},d^c_{iL}$
($\tilde{u}_{iL},\tilde{d}^c_{iL}$ with $i(j)=1,2,3$).

The supersymmetric partner of the scalar Higgs fields, the
higgsinos, are
\begin{eqnarray}
\tilde{ \eta} &=&
\left( \begin{array}{c} \tilde{ \eta}^{0}_{1} \\
                        \tilde{ \eta}^{-} \\
                        \tilde{ \eta}^{0}_{2}
\end{array} \right),\;
\tilde{ \chi} =
\left( \begin{array}{c} \tilde{ \chi}^{0}_{1} \\
                        \tilde{ \chi}^{-} \\
                        \tilde{ \chi}^{0}_{2}
\end{array} \right) \sim ( {\bf 1},{\bf 3},-1/3 ),
\nonumber
\\
\tilde{ \rho} &=&
\left( \begin{array}{c} \tilde{ \rho}^{+}_{1} \\
                        \tilde{ \rho}^{0} \\
                        \tilde{ \rho}^{+}_{2}
\end{array} \right) \sim ( {\bf 1},{\bf 3},2/3 ),
\label{esca}
\end{eqnarray}
and the respective extra higgsinos, needed to cancel the chiral
anomaly of the higgsinos in Eq.~(\ref{esca}), are
\begin{eqnarray}
\tilde{\eta}^{\prime} &=&
\left( \begin{array}{c} \tilde{\eta}^{\prime 0}_{1} \\
                        \tilde{\eta}^{\prime +} \\
                        \tilde{\eta}^{\prime 0}_{2}
         \end{array} \right),
\tilde{\chi}^{\prime} =
\left( \begin{array}{c} \tilde{ \chi}^{\prime 0}_{1} \\
                        \tilde{ \chi}^{\prime +} \\
                        \tilde{ \chi}^{\prime 0}_{2}
\end{array} \right) \sim ( {\bf 1},{\bf 3^{*}},1/3 ),
\nonumber \\
\tilde{\rho}^{\prime} &=&
\left( \begin{array}{c} \tilde{\rho}^{\prime -}_{1} \\
                        \tilde{\rho}^{\prime 0} \\
                        \tilde{\rho}^{\prime -}_{2}
\end{array} \right) \sim ( {\bf 1},{\bf 3^{*}},-2/3 ),
\label{escac}
\end{eqnarray}
and the corresponding scalar partners denoted by
$\eta^{\prime}$,$\chi^{\prime}$, $\rho^{\prime}$, with the same
charge assignment as in Eq.~(\ref{escac}), and with the following
VEVs: $v^{\prime}=\langle \eta^{\prime 0}_1 \rangle$,
$w^{\prime}=\langle \chi^{\prime 0}_2\rangle $ and $
u^{\prime}=\langle \rho^{\prime 0} \rangle$.

Concerning the gauge bosons and their superpartners, if we denote
the gluons by $g^b$ the respective superparticles, the gluinos,
are denoted by $\lambda^b_{C}$, with $b=1, \ldots,8$; and in the
electroweak sector we have $V^b$, the gauge boson of $SU(3)_{L}$,
and their gauginos partners $\lambda^b_{A}$; finally we have the
gauge boson of $U(1)_{N}$, denoted by $V^{\prime}$, and its
supersymmetric partner $\lambda_{B}$. This is the total number of
fields in the minimal supersymmetric extension of the 3-3-1 model
of Refs.~\cite{331rh}.

\section {The supersymmetric scalar potential and mass spectrum}
\label{sec:mass}

It is well known that in supersymmetric models, that the
contributions to the scalar potential arise from three sources -
the auxiliary $F$- and $D$- fields \cite{wb} and the soft terms
\cite{haber}.

On this article we will write only necessary terms to pick all the
terms needed to construct the scalar potential of our model, see
\cite{s331r} to the complete lagrangian.

\subsection{Elimination of the auxiliary fields}

The lagrangian of the gauge sector is a source of the $D$- terms
and it is written as
\begin{eqnarray}
{\cal L}^{gauge}
&=& \frac{1}{4} \int  d^{2}\theta\;Tr[ {\cal W_{C}} {\cal W_{C}}]+
\frac{1}{4} \int  d^{2}\theta\;Tr[ {\cal W_{L}} {\cal W_{L}}]+
\frac{1}{4} \int  d^{2}\theta
{\cal W^{ \prime}}{\cal W^{ \prime}} \nonumber \\
&+&
\frac{1}{4} \int  d^{2}\bar{\theta}\;Tr[
\bar{\cal{W_{C}}}\bar{\cal{W_{C}}}]+ \frac{1}{4}
\int  d^{2}\bar{\theta}\;Tr[\bar{\cal{W_{L}}}\bar{\cal{W_{L}}}]+
\frac{1}{4} \int  d^{2}\bar{\theta}
\bar{\cal{W^{ \prime}}}\bar{\cal{W^{ \prime}}}\,\ ,
\label{gaug}
\end{eqnarray}
where $\cal{W}_{C}$, $\cal{W}$ and $\cal{W}^{ \prime}$ are fields
that can be written as follows
\begin{eqnarray}
\cal{W}_{\zeta C}&=&- \frac{1}{8g_s} \bar{D} \bar{D} e^{-2g_s \hat{V}_{C}}
D_{\zeta} e^{2g_s \hat{V}_{C}},\nonumber \\
\cal{W}_{\zeta L}&=&- \frac{1}{8g} \bar{D} \bar{D} e^{-2g \hat{V}}
D_{\zeta} e^{2g \hat{V}}, \nonumber \\
\cal{W}^{\prime}_{\zeta}&=&- \frac{1}{4} \bar{D} \bar{D} D_{\zeta}
\hat{V}^{\prime}, \,\ \zeta=1,2.
\label{cforca}
\end{eqnarray}
The coupling $g_{s}$ is the gauge coupling constants of $SU(3)_c$
while $g$ and $g^{\prime}$ are the gauge coupling constants of
$SU(3)_L$ and $U(1)_N$, respectively.

In the scalar sector,  both $F$- and $D$- terms yield the
following lagrangian
\begin{eqnarray}
{\cal L}^{scalar}
&=& \int d^{4}\theta\;\left[\,
\hat{ \bar{ \eta}}e^{[2g\hat{V}+g^{\prime} \left( - \frac{1}{3}\right)
\hat{V}^{\prime}]} \hat{ \eta} +
\hat{ \bar{ \chi}}e^{[2g\hat{V}+g^{\prime} \left( - \frac{1}{3}\right)
\hat{V}^{\prime}]} \hat{ \chi} +
\hat{ \bar{ \rho}}e^{[2g\hat{V}+g^{\prime} \left( \frac{2}{3}\right)
\hat{V}^{\prime}]} \hat{ \rho} \right. \nonumber \\
&+& \left. \,
\hat{ \bar{ \eta}}^{\prime}
e^{[2g\hat{ \bar{V}}+g^{\prime} \left( \frac{1}{3}\right) \hat{V}^{\prime}]}
\hat{ \eta}^{\prime}+
\hat{ \bar{ \chi}}^{\prime}
e^{[2g\hat{ \bar{V}}+g^{\prime} \left( \frac{1}{3}\right) \hat{V}^{\prime}]}
\hat{ \chi}^{\prime} +
\hat{ \bar{ \rho}}^{\prime}
e^{[2g\hat{ \bar{V}}+g^{\prime} \left( - \frac{2}{3}\right) \hat{V}^{\prime}]}
\hat{ \rho}^{\prime} \right] \nonumber \\
&+& \int d^{2}\theta \;W+ \int d^{2}\bar{ \theta}\; \overline{W} \!.
\hspace{2mm}
\label{esc}
\end{eqnarray}
$W$ is the superpotential of the model, it is only $F$- term
source. The superpotential is decomposed as follows
\begin{equation}
W= \frac{W_{2}}{2}+ \frac{W_{3}}{3}
\end{equation}
and it can be written explicitly as
\begin{eqnarray}
W_{2}&=&\mu_{0a}\hat{L}_{a} \hat{ \eta}^{\prime}+
\mu_{1a}\hat{L}_{a} \hat{ \chi}^{\prime}+
\mu_{ \eta} \hat{ \eta} \hat{ \eta}^{\prime}+
\mu_{ \chi} \hat{ \chi} \hat{ \chi}^{\prime}+
\mu_{ \rho} \hat{ \rho} \hat{ \rho}^{\prime}, \nonumber \\
W_{3}&=& \lambda_{1ab} \hat{L}_{a} \hat{ \rho}^{\prime} \hat{l}^{c}_{b}+
\lambda_{2a} \epsilon \hat{L}_{a} \hat{\chi} \hat{\rho}+
\lambda_{3a} \epsilon \hat{L}_{a} \hat{\eta} \hat{\rho}+
\lambda_{4ab} \epsilon \hat{L}_{a} \hat{L}_{b} \hat{\rho}+
\kappa_{1i} \hat{Q}_{3} \hat{\eta}^{\prime} \hat{u}^{c}_{i}+
\kappa_{1}^{\prime} \hat{Q}_{3} \hat{\eta}^{\prime} \hat{u}^{\prime c}
\nonumber \\  &+&
\kappa_{2i} \hat{Q}_{3} \hat{\chi}^{\prime} \hat{u}^{c}_{i}+
\kappa_{2}^{\prime} \hat{Q}_{3} \hat{\chi}^{\prime} \hat{u}^{\prime c}+
\kappa_{3\alpha i} \hat{Q}_{\alpha} \hat{\eta} \hat{d}^{c}_{i}+
\kappa_{3\alpha \beta}^{\prime} \hat{Q}_{\alpha}
\hat{\eta} \hat{d}^{\prime c}_{\beta}+
\kappa_{4\alpha i} \hat{Q}_{\alpha}\hat{\rho}\hat{u}^{c}_{i}+
\kappa_{4\alpha}^{\prime} \hat{Q}_{\alpha}\hat{\rho}
\hat{u}^{\prime c} \nonumber \\
&+&
\kappa_{5i}\hat{Q}_{3} \hat{\rho}^{\prime} \hat{d}^{c}_{i}+
\kappa_{5 \beta}^{\prime}\hat{Q}_{3} \hat{\rho}^{\prime} \hat{d}^{c}_{\beta}+
\kappa_{6\alpha i} \hat{Q}_{\alpha} \hat{\chi} \hat{d}^{c}_{i}+
\kappa_{6\alpha \beta}^{\prime} \hat{Q}_{\alpha}
\hat{\chi} \hat{d}^{\prime c}_{\beta}+
f_{1} \epsilon \hat{ \rho} \hat{ \chi} \hat{ \eta}+
f^{\prime}_{1} \epsilon \hat{ \rho}^{\prime}
\hat{ \chi}^{\prime}\hat{ \eta}^{\prime} \nonumber \\
&+&
\zeta_{\alpha \beta \gamma} \epsilon \hat{Q}_{\alpha}
\hat{Q}_{\beta} \hat{Q}_{\gamma}+
\lambda^{\prime}_{\alpha ai}\hat{Q}_{\alpha}\hat{L}_{a} \hat{d}^{c}_{i}+
\lambda^{\prime \prime}_{ijk} \hat{d}^{c}_{i} \hat{u}^{c}_{j} \hat{d}^{c}_{k}+
\xi_{1ij \beta} \hat{d}^{c}_{i} \hat{u}^{c}_{j} \hat{d}^{\prime c}_{\beta}+
\xi_{2 \alpha a \beta}\hat{Q}_{\alpha}
\hat{L}_{a} \hat{d}^{\prime c}_{\beta} \nonumber \\
&+&
\xi_{3i \beta} \hat{d}^{c}_{i} \hat{u}^{\prime c} \hat{d}^{\prime c}_{\beta}+
\xi_{4ij} \hat{d}^{c}_{i} \hat{u}^{\prime c} \hat{d}^{c}_{j}+
\xi_{5 \alpha i \beta} \hat{d}^{\prime c}_{\alpha}
\hat{u}^{c}_{i} \hat{d}^{\prime c}_{\beta}+
\xi_{6 \alpha \beta} \hat{d}^{\prime c}_{\alpha}
\hat{u}^{\prime c} \hat{d}^{\prime c}_{\beta}
\label{sp3susy2}
\end{eqnarray}
The coefficients $\mu_{0}, \mu_{1}, \mu_{\eta}, \mu_{\rho}$ and $\mu_{\chi}$
have mass dimension, while all the coefficients in $W_{3}$ are dimensionless.

To get the scalar potential of our model we have to pick up the
$F$ and $D$- terms, from  Eqs.
(\ref{gaug},\ref{esc},\ref{sp3susy2}) and get
\begin{eqnarray}
{\cal L}_{F}&=&{\cal L}^{scalar}_{F}+{\cal L}^{W2}_{F}+{\cal L}^{W3}_{F}
\nonumber \\
&=&\vert F_{\eta} \vert^2+ \vert F_{\rho} \vert^2+
\vert F_{\chi} \vert^2+
\vert F_{\eta^{\prime}} \vert^2+
\vert F_{\rho^{\prime}} \vert^2+
\vert F_{\chi^{\prime}} \vert^2 \nonumber \\
&+& \frac{\mu_{ \eta}}{2}( \eta F_{\eta^{\prime}}+
\eta^{\prime} F_{ \eta}+ \eta^{\dagger} F^{\dagger}_{\eta^{\prime}}+
\eta^{\prime \dagger} F^{\dagger}_{ \eta})+
\frac{\mu_{ \rho}}{2}( \rho F_{\rho^{\prime}}+ \rho^{\prime} F_{ \rho}+
\rho^{\dagger} F^{\dagger}_{\rho^{\prime}}+ \rho^{\prime \dagger}
F^{\dagger}_{ \rho})
\nonumber \\
&+&
\frac{\mu_{ \rho}}{2}( \rho F_{\rho^{\prime}}+ \rho^{\prime} F_{ \rho}+
\rho^{\dagger} F^{\dagger}_{\rho^{\prime}}+ \rho^{\prime \dagger}
F^{\dagger}_{ \rho})
+ \frac{1}{3} \left[ f_{1} \epsilon (F_{ \rho} \chi \eta+
\rho F_{ \chi} \eta+ \rho \chi F_{ \eta} \right. \nonumber \\
&+& \left. F^{\dagger}_{ \rho} \chi^{\dagger} \eta^{\dagger}+
\rho^{\dagger} F^{\dagger}_{ \chi} \eta^{\dagger}+
\rho^{\dagger} \chi^{\dagger} F^{\dagger}_{ \eta}) + f^{\prime}_{1} \epsilon (
F_{ \rho^{\prime}} \chi^{\prime} \eta^{\prime}+
\rho^{\prime} F_{ \chi^{\prime}} \eta^{\prime}+
\rho^{\prime} \chi^{\prime} F_{ \eta^{\prime}} \right. \nonumber \\
&+& \left. F^{\dagger}_{ \rho^{\prime}} \chi^{\prime \dagger}
\eta^{\prime \dagger}+
\rho^{\prime \dagger} F^{\dagger}_{ \chi^{\prime}} \eta^{\prime \dagger}+
\rho^{\prime \dagger} \chi^{\prime \dagger} F^{\dagger}_{ \eta^{\prime}})
 \right] \,\ , \nonumber \\
{\cal L}_{D}&=&{\cal L}^{gauge}_{D}+{\cal L}^{scalar}_{D} \nonumber \\
&=& \frac{1}{2}D^{a}D^{a}+ \frac{1}{2}DD
+ \frac{g}{2} \left[ \eta^{\dagger}\lambda^a\eta+
\rho^{\dagger}\lambda^a\rho+ \chi^{\dagger}\lambda^a\chi-
\eta^{\prime \dagger}\lambda^{* a}\eta^{\prime}-
\rho^{\prime \dagger}\lambda^{* a}\rho^{\prime} \right. \nonumber \\
&-& \left.
\chi^{\prime \dagger}\lambda^{* a}\chi^{\prime} \right] D^{a}+
\frac{g^{ \prime}}{2} \left[ - \frac{1}{3} \eta^{\dagger} \eta +
\frac{1}{3} \eta^{\prime \dagger} \eta^{\prime}-
\frac{1}{3} \chi^{\dagger} \chi +
\frac{1}{3} \chi^{\prime \dagger} \chi^{\prime}+
\frac{2}{3} \rho^{\dagger}\rho
- \frac{2}{3} \rho^{\prime \dagger}\rho^{\prime}
 \right]D.
\label{auxiliarm1}
\end{eqnarray}

We will now show that these fields can be eliminated through the
Euler-Lagrange equations
\begin{eqnarray}
\frac{\partial {\cal L}}{\partial \phi}- \partial_{m}
\frac{\partial {\cal L}}{\partial (\partial_{m} \phi)}=0 \,\ ,
\label{Euler-Lagrange Equation}
\end{eqnarray}
where $\phi = \eta , \rho , \chi , \eta^{\prime}, \rho^{\prime},
\chi^{\prime}$. Formally auxiliary fields are defined as fields
having no kinetic terms. Thus, this definition immediately yields
that the Euler-Lagrange equations for auxiliary fields simplify to
$\frac{\partial {\cal L}}{\partial \phi}=0$.

Applying these simplified equations to various auxiliary $F$-fields yields
the following relations
\begin{eqnarray}
F^{\dagger}_{\eta}&=&- \left( \frac{\mu_{\eta}}{2}\eta^{\prime}+
\frac{f_{1}}{3} \epsilon \rho \chi \right) \,\ ;
\,\ F_{\eta}=- \left( \frac{\mu_{\eta}}{2}\eta^{\prime \dagger}+
\frac{f_{1}}{3} \epsilon \rho^{\dagger} \chi^{\dagger} \right) \,\ ,
\nonumber \\
F^{\dagger}_{\rho}&=&- \left( \frac{\mu_{\rho}}{2}\rho^{\prime}+
\frac{f_{1}}{3} \epsilon \chi \eta \right) \,\ ;
\,\ F_{\rho}=- \left( \frac{\mu_{\rho}}{2}\rho^{\prime \dagger}+
\frac{f_{1}}{3} \epsilon \chi^{\dagger} \eta^{\dagger} \right) \,\ ,
\nonumber \\
F^{\dagger}_{\chi}&=&- \left( \frac{\mu_{\chi}}{2}\chi^{\prime}+
\frac{f_{1}}{3} \epsilon \rho \eta \right) \,\ ;
\,\ F_{\chi}=- \left( \frac{\mu_{\chi}}{2}\chi^{\prime \dagger}+
\frac{f_{1}}{3} \epsilon \rho^{\dagger} \eta^{\dagger} \right) \,\ ,
\nonumber \\
F^{\dagger}_{\eta^{\prime}}&=&- \left( \frac{\mu_{\eta}}{2}\eta +
\frac{f^{\prime}_{1}}{3} \epsilon \rho^{\prime}\chi^{\prime}
\right) \,\ ;
\,\  F_{\eta^{\prime}}=- \left( \frac{\mu_{\eta}}{2}\eta^{\dagger}+
\frac{f^{\prime}_{1}}{3} \epsilon \rho^{\prime \dagger}\chi^{\prime \dagger}
\right) \,\ , \nonumber \\
F^{\dagger}_{\rho^{\prime}}&=&- \left( \frac{\mu_{\rho}}{2}\rho +
\frac{f^{\prime}_{1}}{3} \epsilon \chi^{\prime}\eta^{\prime}
\right) \,\ ;
\,\ F_{\rho^{\prime}}=- \left( \frac{\mu_{\rho}}{2}\rho^{\dagger}+
\frac{f^{\prime}_{1}}{3} \epsilon \chi^{\prime \dagger}\eta^{\prime \dagger}
\right) \,\ , \nonumber \\
F^{\dagger}_{\chi^{\prime}}&=&- \left( \frac{\mu_{\chi}}{2}\chi +
\frac{f^{\prime}_{1}}{3} \epsilon \rho^{\prime} \eta^{\prime}
\right) \,\ ; \,\  F_{\chi^{\prime}}=- \left(
\frac{\mu_{\chi}}{2}\chi^{\dagger}+ \frac{f^{\prime}_{1}}{3}
\epsilon \rho^{\prime \dagger}\eta^{\prime \dagger} \right) \,\ .
\end{eqnarray}
Using these equations, we can rewrite Eq.(\ref{auxiliarm1}) as
\begin{equation}
{\cal L}_{F}=- \left(
\vert F_{\eta} \vert^2+ \vert F_{\rho} \vert^2+ \vert F_{\chi} \vert^2+
\vert F_{\eta^{\prime}} \vert^2+ \vert F_{\rho^{\prime}} \vert^2+
\vert F_{\chi^{\prime}} \vert^2 \right) .
\end{equation}
 Performing the same program to $D$-fields we get
\begin{eqnarray}
D^{a}&=&- \frac{g}{2} \left[ \eta^{\dagger}\lambda^a\eta+
\rho^{\dagger}\lambda^a\rho+ \chi^{\dagger}\lambda^a\chi-
\eta^{\prime \dagger}\lambda^{* a}\eta^{\prime}-
\rho^{\prime \dagger}\lambda^{* a}\rho^{\prime}-
\chi^{\prime \dagger}\lambda^{* a}\chi^{\prime} \right] \,\ , \nonumber \\
D&=&- \frac{g^{ \prime}}{2} \left[ - \frac{1}{3} \eta^{\dagger} \eta +
\frac{1}{3} \eta^{\prime \dagger} \eta^{\prime}-
\frac{1}{3} \chi^{\dagger} \chi +
\frac{1}{3} \chi^{\prime \dagger} \chi^{\prime}+
\frac{2}{3} \rho^{\dagger}\rho
- \frac{2}{3} \rho^{\prime \dagger}\rho^{\prime}
 \right] \,\ ,
\end{eqnarray}
which is in accordance with Eq.(\ref{auxiliarm1})
\begin{equation}
{\cal L}_{D}=- \frac{1}{2} \left( D^{a}D^{a}+DD
\right).
\end{equation}

\subsection{The soft term}

Now we are considering  the last source to construct our scalar
potential. The most general soft supersymmetry breaking terms,
which do not induce quadratic divergence, where described by
Girardello and Grisaru \cite{10}. They found that the allowed
terms can be categorized as follows: a scalar field $A$ with mass
terms
\begin{equation}
-m^{2} A^{\dagger}A,
\end{equation}
 a fermion field gaugino  $\lambda$ with mass  terms
\begin{equation}
- \frac{1}{2} (M_{ \lambda} \lambda^{a} \lambda^{a}+H.c)
\end{equation}
and finally trilinear scalar interaction terms
\begin{equation}
\epsilon^{ijk}A_{i}A_{j}A_{k}+H.c.
\end{equation}
The terms on this case are similar with the terms allowed in the
superpotential of the model we are considering.

Of course, the form of these terms, depend on the model we are
considering, in our case see \cite{s331r}. The only necessary part
 for us on this article is given by
\begin{eqnarray}
{\cal L}^{\mbox{soft}}_{\mbox{scalar}}&=& -m^2_{ \eta}\eta^{
\dagger}\eta-m^2_{ \rho}\rho^{ \dagger}\rho- m^2_{ \chi}\chi^{
\dagger}\chi - m^2_{\eta^{\prime}}\eta^{\prime
\dagger}\eta^{\prime}- m^2_{\rho^{\prime}}\rho^{\prime
\dagger}\rho^{\prime}- m^2_{\chi^{\prime}}\chi^{\prime
\dagger}\chi^{\prime}
\nonumber \\
&+&[k_1\epsilon_{ijk}\rho_i\chi_j\eta_k+
k^{\prime}_1\epsilon_{ijk}\rho^{\prime}_i\chi^{\prime}_j\eta^{\prime}_k+H.c.].
\label{potencial}
\end{eqnarray}

\section{The supersymmetric scalar potential and mass spectrum}
\label{sec:massspectrum}

The pattern of the symmetry breaking in this  model is given by
the following scheme
\begin{eqnarray}
&\mbox{MSUSY331}&
\stackrel{{\cal L}_{soft}}{\longmapsto}
\mbox{SU(3)}_C\ \otimes \ \mbox{SU(3)}_L\otimes \mbox{U(1)}_N
\stackrel{\langle\chi\rangle \langle \chi^{\prime}\rangle}{\longmapsto}
\mbox{SU(3)}_C \ \otimes \ \mbox{SU(2)}_L\otimes
\mbox{U(1)}_Y \nonumber \\
&\stackrel{\langle\rho,\eta, \rho^{\prime}\eta^{\prime}\rangle}{\longmapsto}&
\mbox{SU(3)}_C \ \otimes \ \mbox{U(1)}_Q
\end{eqnarray}

 In the considered  model, we  have three extra triplets, therefore in the neutral
 scalar sector we have $10\times10$ matrix instead of $5\times5$ in the
non-supersymmetric gauge theory $SU_c(3)\bigotimes
SU_L(3)\bigotimes U_N(1)$ \cite{longlan,l97}. As  mentioned in
~\cite{s331r}, the supersymmetric Higgs potential can be written:
 \be V_{331SUSYRN} = V_F +
V_D + V_{\rm soft}, \label{p1} \ee where \bea V_F & = & -{\cal
L}_F = \sum_m F^\dagger_m  F_m \crn & =
&\sum_{ijk}\left[\left\vert\frac{\mu_\eta}{2}\eta^\prime_i+
\frac{f_1}{3}\epsilon_{ijk}\rho_j\chi_k \right\vert^2+ \left\vert
\frac{\mu_\chi}{2}\chi^\prime_i+\frac{f_1}{3}\epsilon_{ijk}\eta_j\rho_k
\right\vert^2+ \left\vert
\frac{\mu_\rho}{2}\rho^\prime_i+\frac{f_1}{3}\epsilon_{ijk}\chi_j\eta_k
\right\vert^2\right.\crn & + &\left.\left\vert
\frac{\mu_\eta}{2}\eta_i+\frac{f^\prime_1}{3}\epsilon_{ijk}\rho^\prime_j
\chi^\prime_k\right\vert^2+ \left\vert
\frac{\mu_\chi}{2}\chi_i+\frac{f^\prime_1}{3}\epsilon_{ijk}
\eta^\prime_j\rho^\prime_k\right\vert^2+ \left\vert
\frac{\mu_\rho}{2}\rho_i+\frac{f^\prime_1}{3}\epsilon_{ijk}
\chi^\prime_j\eta^\prime_k\right\vert^2\right] \label{p2} \eea The
second term is given by\bea V_D&=&-{\cal
L}_D=\frac{1}{2}(D^aD^a+DD)=\frac{g^{\prime2}}{2} \left( -
\frac{1}{3} \eta^{\dagger} \eta + \frac{1}{3} \eta^{\prime
\dagger} \eta^{\prime}- \frac{1}{3} \chi^{\dagger} \chi +
\frac{1}{3} \chi^{\prime \dagger} \chi^{\prime}+ \frac{2}{3}
\rho^{\dagger}\rho - \frac{2}{3} \rho^{\prime
\dagger}\rho^{\prime} \right)^2\crn & +
&\frac{g^2}{8}(\eta^\dagger_i\lambda^b_{ij}\eta_j-
\eta^{\prime\dagger}_i\lambda^{*b}_{ij}\eta^\prime_j+
\chi^\dagger_i\lambda^b_{ij}\chi_j-
\chi^{\prime\dagger}_i\lambda^{*b}_{ij}\chi^\prime_j+
\rho^\dagger_i\lambda^b_{ij}\rho_j-
\rho^{\prime\dagger}_i\lambda^{*b}_{ij}\rho^\prime_j)^2\!,
\label{p3} \eea
 and
\bea V_{\rm soft}&=&-{\cal L}_{\rm soft}=
m^2_\eta\eta^\dagger\eta+m^2_\rho\rho^\dagger\rho+m^2_\chi\chi^\dagger\chi
+m^2_{\eta^\prime}\eta^{\prime\dagger}\eta^\prime \crn &+&
m^2_{\rho^\prime}\rho^{\prime\dagger}\rho^\prime+
m^2_{\chi^\prime}\chi^{\prime\dagger}\chi^\prime-
\epsilon_{ijk}(k_1 \rho_i\chi_j\eta_k+
k^{\prime}_1\rho^\prime_i\chi^\prime_j\eta^\prime_k +  h.c.). \label{p4}
\eea

\hs For convenience, we rewrite the expansion of the neutral
scalar fields:
 \bea
\eta &= & \left(%
\begin{array}{c}
   v+\eta_1 + i\eta_2 \\
  0 \\
  \eta_3 + i\eta_4
\end{array}%
\right),\hs
 \chi  =  \left(%
 \begin{array}{c}
   \chi_1 + i\chi_2 \\  0 \\ w + \chi_3 + i\chi_4
 \end{array}%
 \right),\hs
 \rho =  \left(%
 \begin{array}{c}
  0\\ u + \rho_1 + i\rho_2 \\ 0
  \end{array}%
\right)
 \label{p5}
\eea

Similarly for the prime fields \bea
 \eta^\prime &= & \left(%
\begin{array}{c}
v^\prime+\eta^\prime_1 + i\eta^\prime_2\\ 0 \\
\eta^\prime_3 + i\eta^\prime_4
\end{array}%
\right),\hs
 \chi^\prime  =  \left(%
 \begin{array}{c}
\chi^\prime_1 + i\chi^\prime_2 \\
 0 \\ w^\prime + \chi^\prime_3 + i\chi^\prime_4
 \end{array}%
 \right),\hs
 \rho^\prime  =  \left(%
 \begin{array}{c}
0\\ u^\prime + \rho^\prime_1 + i\rho^\prime_2 \\ 0
  \end{array}%
\right)
 \label{p6}
\eea
For the sake of simplicity, here we assume that
vacuum expectation values (VEVs) are real. This means that
the CP violation through the scalar exchange is not considered
in this work.

Then, the real part, of Eqs.(\ref{p5},\ref{p6}),
($\eta_1,  \eta'_1, \chi_1, \chi'_1,...$) are
called CP-even scalar or scalar, while imaginary parts ($\eta_2,
\eta'_2, \chi_2, \chi'_2,...$) - CP-odd scalar or pseudoscalar
field. In this paper, we call them scalar and pseudoscalar.
 Returning to Eqs (\ref {p2})-(\ref{p4}), by  requirement of vanishing
 the linear terms in fields, we get,  in the tree level approximation, the
following constraint equations
 \bea
m_{\eta}^2&=&-\frac{1}{36v}\left(- 6vw^2g^2 +4vw^2f_1^2
+4vw^2g^{\prime2} - 6vu^2g^2 + 4vu^2f_1^2 -
8vu^2g^{\prime2}\right.\crn & -&
 12vv^{\prime^2}g^2 -4vv^{\prime^2}g^{\prime2} -
 4vw^{\prime2}g^{\prime2} + 6vu^{\prime^2}g^2 - 36wuk_1
+ 8vu^{\prime^2}g^{\prime2}\crn &+&\left. 9v\mu_{\eta}^2 +
12v^3g^2 +2v^3g^{\prime2}  + 6wu^{\prime}\mu_{\rho}f_1 +
6uw^{\prime}\mu_{\chi}f_1 +
6w^{\prime}u^{\prime}\mu_{\eta}f^{\prime}_1\right),\crn
 m_{\chi}^2
&=&-\frac{1}{36w}\left(- 36vuk_1 + 6vu^{\prime}\mu_{\rho}f_1 -
6v^2wg^2
       +4v^2wf_1^2 +4v^2w
 g^{\prime2}+ 6v^{\prime}u^{\prime}\mu_{\chi}f^{\prime}_1
 \right.\crn &-& 6wu^2g^2 +4wu^2f_1^2
- 8wu^2g^{\prime2} + 6w
 v^{\prime^2}g^2 -4wv^{\prime^2}g^{\prime2}+6uv^{\prime}\mu_{\eta}f_1
 \crn &-& \left.12ww^{\prime2}g^2 -4ww^{\prime2}g^{\prime2} +
 6wu^{\prime^2}g^2 + 8wu^{\prime^2}g^{\prime2}
 + 9w\mu_{\chi}^2 + 12w^3g^2 +4
   w^3g^{\prime2}\right),\crn
 m_{\rho}^2&=&-\frac{1}{36u}\left(- 36vwk_1 +
6vw^{\prime}\mu_{\chi}f_1
 - 6v^2ug^2 +4v^2uf_1^2 - 8v^2u
 g^{\prime2} +6wv^{\prime}\mu_{\eta}f_1\right.
\crn &-& 6w^2ug^2 +4w^2uf_1^2 -8w^2u
 g^{\prime2} + 6uv^{\prime^2}g^2 +
8uv^{\prime^2}g^{\prime2} + 6uw^{\prime2}g^2 + 16u^3g^{\prime2}
\crn &+ &\left.8uw^{\prime2}g^{\prime2}
 - 12uu^{\prime^2}g^2 - 16uu^{\prime^2}g^{\prime2}
 + 9u\mu_{\rho}^2 + 12u^3g^2+
6v^{\prime}w^{\prime}\mu_{\rho}f^{\prime}_1\right),\crn
  m_{\eta^\prime}^2&=&-\frac{1}{ 36v^{\prime}}\left(-
12v^2v^{\prime}g^2 -4v^2v^{\prime}g^{\prime2}
 + 6wu\mu_{\eta}f_1 + 6wu^{\prime}\mu_{\chi}f^{\prime}_1 +
 6w^2v^{\prime}g^2 \right. \crn &-&4w^2v^{\prime}g^{\prime2}
 + 6uw^{\prime}\mu_{\rho}f^{\prime}_1 + 6u^2v^{\prime}g^2
 + 8u^2v^{\prime}g^{\prime2} - 6v^{\prime}w^{\prime2}g^2
+4v^{\prime}w^{\prime2}(f^{\prime}_1)^2 +4v^{\prime}
 w^{\prime2}g^{\prime2}\crn &-& \left.36 w^{\prime}u^{\prime}k^{\prime}_1
 - 6v^{\prime}u^{\prime^2}g^2
+4v^{\prime}u^{\prime^2}(f^{\prime}_1)^2
  - 8v^{\prime}u^{\prime^2}g^{\prime2}
 + 9v^{\prime}\mu_{\eta}^2 + 12v^{\prime^3}g^2
 +4v^{\prime^3}g^{\prime2} \right),\crn
 m_{\chi^\prime}^2 &=&
-\frac{1}{36w^{\prime}}\left(6vu\mu_{\chi}f_1 +
6vu^{\prime}\mu_{\eta}f^{\prime}_1 -4v^2w^{\prime}g^{\prime2} -
12w^2w^{\prime}g^2 - 4w^2w^{\prime}g^{\prime2} +
6uv^{\prime}\mu_{\rho}f^{\prime}_1 \right. \crn &+&
6u^2w^{\prime}g^2 + 8u^2w^{\prime} g^{\prime2} -
36v^{\prime}u^{\prime}k^{\prime}_1 - 6v^{\prime^2}w^{\prime}g^2
+4v^{\prime^2}w^{\prime}(f^\prime_1)^2 +4v^{\prime^2}
w^{\prime}g^{\prime2} \crn &-& \left.6w^{\prime}u^{\prime^2}g^2
         +4w^{\prime}u^{\prime^2}(f^\prime_1)^2
         - 8w^{\prime}u^{\prime^2}g^{\prime2} +
         9w^{\prime}\mu_{\chi}^2 + 12w^{\prime3}g^2
         +4w^{\prime3}g^{\prime2}\right),\crn
m_{\rho^\prime}^2&=-&\frac{1}{36u^{\prime}}\left( 6vw\mu_{\rho}f_1
+ 6vw^{\prime}\mu_{\eta}f^{\prime}_1 + 6v^2u^{\prime}g^2 +
8v^2u^{\prime}g^{\prime2} + 6wv^{\prime}\mu_{\chi}f^{\prime}_1 +
6w^2u^{\prime}g^2 \right.\crn & +& 8w^2u^{\prime}g^{\prime2} -
12u^2u^{\prime}g^2
          - 16u^2u^{\prime}g^{\prime2} - 36v^{\prime}w^{\prime}k^{\prime}_1
- 6v^{\prime^2}u^{\prime}g^2 +4v^{\prime^2}u^{\prime}
(f^\prime_1)^2- 8v^{\prime^2}u^{\prime}g^{\prime2}
         \crn &-&\left. 6w^{\prime2}u^{\prime}g^2
         +4w^{\prime2}u^{\prime}(f^\prime_1)^2 - 8
         w^{\prime2}u^{\prime}g^{\prime2} + 9u^{\prime}\mu_{\rho}^2
         + 12u^{\prime^3}g^2 +
         16u^{\prime^3}g^{\prime2}\right).
 \eea

The most general scalar potential (i.e. the one including all
terms consistent with the gauge invariance and renormalizability
as considered in \cite{s331r}) is very complicated. However, we
can use one approximation to simplify our analysis. Before writing
this approximation, we will analyze the gauge sector.

We can write the gauge mass term as
\begin{eqnarray}
{\cal L}_{Higgs}&=& ({\cal D}_{m} \eta)^{\dagger}({\cal D}^{m} \eta)+
({\cal D}_{m} \rho)^{\dagger}({\cal D}^{m} \rho)+
({\cal D}_{m} \chi)^{\dagger}({\cal D}^{m} \chi)+
(\overline{{\cal D}_{m}} \eta^{\prime})^{\dagger}
(\overline{{\cal D}^{m}} \eta^{\prime}) \nonumber \\ &+&
(\overline{{\cal D}_{m}} \rho^{\prime})^{\dagger}(\overline{{\cal D}^{m}} \rho^{\prime})+
(\overline{{\cal D}_{m}} \chi^{\prime})^{\dagger}(\overline{{\cal D}^{m}} \chi^{\prime}),
\end{eqnarray}
where ${\cal D}_{m}$ is the triplet covariant derivative given by
\begin{equation}
{\cal D}_{m}\phi_{i}  =  \partial_{m}\phi_{i} - ig\left( \vec{V}_{m}
.\frac{\vec{\lambda}}{2}\right)^{j}_{i}\phi_{j} -
ig^\prime N_{\phi_{i}}V^{\prime}_{m}\phi_{i},
\end{equation}
while $\overline{{\cal D}_{m}}$ is the anti-triplet covariant derivative, and
it is written as
\begin{equation}
\overline{{\cal D}_{m}}\phi_{i}  =  \partial_{m}\phi_{i} +
ig\left( \vec{V}_{m}
.\frac{\vec{\overline{\lambda}}}{2}\right)^{j}_{i}\phi_{j} +
ig^\prime N_{\phi_{i}}V^{\prime}_{m}\phi_{i},
\end{equation}
and
$\phi = \eta , \rho , \chi , \eta^{\prime} , \rho^{\prime} , \chi^{\prime}$.

The non-Hermitian gauge bosons $\sqrt{2}\ W^{+}_{m}=
V^{1}_{m}-iV^{2}_{m}, \sqrt{2}\ Y^{-}_{m}= V^{6}_{m}-
iV^{7}_{m},\sqrt{2}\ X^{0}_{m}= V^4_{m}- iV^{5}_{m}$  have the
following masses~\cite{331rh}:
\begin{equation}
M^2_W=\frac{g^{2}}{4}(U^2+V^2), M^2_Y=\frac{g^{2}}{4}(U^2+W^2),
M^2_X=\frac{g^{2}}{4}(V^2+W^2) \label{mnhb}
\end{equation}
where  $V^2=v^2+v^{\prime 2}$, $U^2=u^2+u^{\prime 2}$ and
$W^2=w^2+w^{\prime 2}$.

While in the   ($V_{3m},V_{8m},V^{\prime}_{m}$) basis we have the
mass square of the real vector bosons given by:
\begin{equation}
\frac{g^2}{4}\,\left(
\begin{array}{ccc}
V^2+U^2 &\frac{1}{\sqrt3}\left(V^2-U^2 \right)&-\frac{2t}{3}\left(V^2+2U^2\right) \\
\frac{1}{\sqrt3}\left(V^2-U^2 \right)& \frac{1}{3}\left(V^2+U^2+4W^2 \right) &
 -\frac{2t}{3\sqrt3}\left(V^2-2U^2-2W^2 \right) \\
-\frac{2t}{3}\left(V^2+2U^2\right)&-\frac{2t}{3\sqrt3}
\left(V^2-2U^2-2W^2 \right)& \frac{4t^2}{9}\left( V^2+4U^2+W^2\right)\!,
\end{array}
\right)
\label{massneutros}
\end{equation}

The eigenstates of Eq.~(\ref{massneutros}) are
\begin{equation}
A_{m}=\frac{\sqrt3}{4t^2+3}\left(t\,V_{3m}-\frac{t}{\sqrt3}V_{8m}+V^{\prime}_{m} \right),
\label{foton}
\end{equation}
for the photon, and
\begin{equation}
Z^{0}_{m}\approx\frac{3t}{4t^2+15t^2+9}\left[ -\left(\frac{t^2+3}{3t}
\right)V_{3m}-\frac{t}{\sqrt3}V_{8m}+V^{\prime}_{m}\right],
\label{z}
\end{equation}
and
\begin{equation}
Z^{0\prime}_{m}\approx\frac{t}{t^2+3}\left( \frac{\sqrt3}{t}V_{8m}+V^{\prime}_{m}\right)
\label{zp}
\end{equation}
for the $Z^{0}$ and $Z^{0\prime}$, we have neglected the mixing among $Z^0$ and
$Z^{0\prime}$. Their masses are
\begin{eqnarray}
M^{2}_{A}&=&0 \\
M^{2}_{Z}&\simeq& \frac{g^{2}}{4} \left( \frac{3+4t^2}{3+t^2} \right)
(V^2+U^2) \\
M^{2}_{Z^{\prime}}&\simeq& \frac{g^{2}}{3} \left( 1+ \frac{t^2}{3} \right)
W^2.
\end{eqnarray}
so that $M^2_Z/M^2_W\approx(3+4t^2)/(3+t^2)=1/\cos^2\theta_W$, and
\begin{equation}
t^2=\left(\frac{g^\prime}{g}\right)^2=\frac{\sin^2\theta_W}{1-\frac{4}{3}
\sin^2\theta_W}.
\label{sin331}
\end{equation}
Eq. (\ref{mnhb})  yields the splitting between the bilepton masses
\be \vert M_X^2 - M^2_Y \vert  \leq M^2_W \label{massspliting} \ee
which is the same as in the non-supersymmetric version
\cite{longinami}.

The Eqs.(\ref{massneutros},\ref{mnhb}) are very important, because the new
gauge bosons must be sufficiently heavy to keep consistency with low energy
phenomenology. Due this fact the VEV's satisfy the conditions~\cite{s331r}:
\be
w,w^\prime \gg v,v^\prime, u,u^\prime\label{cond}
\ee
 which are followed from vanishing coupling constant of
 singlet fields to the SM gauge bosons.
 In this paper, we will use
 this approximation.

The mass matrices, thus, can be calculated, using
\begin{equation}
M_{ij}^2=\frac{\partial ^2V_{MSUSY331}}{\partial \phi _i\partial \phi _j}
\label{calculomassamodelo1}
\end{equation}
and evaluated at the chosen minimum, where $\phi= \eta , \rho ,
\chi , \eta^{\prime}, \rho^{\prime}, \chi^{\prime}$.

\subsection{Spectrum in the neutral scalar sector}
 In the approximation (\ref{cond}),
 the mass square matrix  of scalar particles in the
 base of $(\eta_1,\rho_1$, $\eta^\prime_1$, $\rho^\prime_1$, $\eta_3,
 \eta^\prime_3$, $\chi_1,\chi^\prime_1$, $\chi_3,\chi^\prime_3)$,
 after imposing the constraint equation, has  the form following

\be
 M_H^2 =\left (\ba{ccccc}
 M_{4\eta\rho}^2
 & 0 & 0 & 0&0 \\ 0 & m_{\eta_3}^2 &0& 0 & 0 \\
 0 & 0 &m_{\eta^\prime_3}^2& 0 & 0\\
 0 & 0 & 0 & M_{2\chi_1\chi_1^\prime}^2 & 0
 \\ 0 & 0 & 0 & 0 &  M_{2\chi_3\chi_3^\prime}^2 \ea \right),
 \label{matr44}
\ee where \be M_{4\eta\rho}^2=\left (\ba{cccc} 0 &
m^2_{\eta_1\rho_1} & 0& m^2_{\eta_1\rho_1^\prime}
\\  m^2_{\eta_1\rho1} & 0 & m^2_{\eta^\prime_1\rho_1}&0 \\ 0&
m^2_{\eta^\prime_1\rho_1}  & 0 & m^2_{\eta^\prime_1\rho^\prime_1}
 \\ m^2_{\eta_1\rho^\prime_1} & 0 &  m^2_{\eta^\prime_1\rho^\prime_1} & 0\ea
 \right),
\label{ma1}\ee
\begin{equation}
M_{2\chi_1\chi^\prime_1}^2=\frac{g^2}{2}\left (\ba{cc} w^{\prime
2} &-w w^{\prime}
\\ -w w^{\prime}&w \ea
\right), \label{ma3}\end{equation}
\begin{equation}
M_{2\chi_3\chi^\prime_3}^2= \frac{2}{9} ( g^{\prime 2} + 3 g^2)
 \left (\ba{cc} w^2 & -w w^{\prime}
\\ -w w^\prime & w^{\prime 2} \ea
\right). \label{ma4}\end{equation} We see that, at the tree level,
$\eta_3, \eta'_3$ are eigenstates {\it already} with masses \bea
 m^2_{\eta_3}& =& \frac{1}{18}w^2\left(9g^2 - 2f_1^2\right),
 \label{dma1}\\
  m^2_{\eta^\prime_3}&=&
   - \frac{1}{2}w^2g^2 + \frac{1}{18}w^{\prime2}\left(9g^2
   - 2\left(f^{\prime}_1\right)^2\right).
 \label{dmas2}
 \eea
We remind that $\eta_3, \eta'_3$, $\chi_1, \chi'_1$ are bilepton,
while $\eta_1, \rho_1, \eta'_1, \rho'_1$, $\chi_3, \chi'_3$ are
pure scalars (without lepton number). From (\ref{matr44}) it
follows that there is no mixing between scalars having different
lepton numbers.

 From (\ref{ma3}) and (\ref{ma4}), it is easily to check that
 \bea
 \mbox{Det} M^2_{\chi_1,\chi_1^{\prime}}&=& \mbox{Det} M^2_{\chi_3,\chi_3^{\prime}}=0, \\
 \mbox{Tr} M^2_{\chi_1,\chi_1^{\prime}}&=
 & \frac{g^2}{2}\left(w^2 +w^{\prime 2}\right),\\
 \mbox{Tr} M^2_{\chi_3,\chi_3^{\prime}}&=&\frac{1}{9} ( g^{\prime 2} + 3 g^2)
 \left(w^2+w'^2\right).\label{dm1}
\eea Therefore, there are two neutral scalar Goldstone bosons and
another massive scalar \bea m^2_{\zeta_{\chi_1
\chi_1^\prime}}&=&\frac{g^2}{2}\left(w^2+w^{\prime
2}\right),\\m^2_{\zeta_{\chi_3 \chi_3^\prime}}&=&\frac{2}{9} (
g^{\prime 2} + 3 g^2)\left(w^2+w'^2\right) \leq
m^2_{\zeta_{\chi_1 \chi_1^\prime}}. \label{ln1}\eea
  \hs Now, we consider $4\times4$ mass matrix $M^2_{4 \eta \rho}$ of
$\eta_1,\rho_1, \eta'_1,\rho'_1$ mixing. The elements of $M^2_{4 \eta \rho}$
given at Eq.(\ref{ma1}) are
\begin{eqnarray}
m^{2}_{\eta_{1}\rho_{1}}&=&\frac{f_{1}}{6}\mu_{\chi}w^{\prime}-k_{1}w
,
\nonumber \\
m^{2}_{\eta_{1}\rho_{1}^{\prime}}&=&\frac{1}{6}\left(
f_{1}\mu_{\rho}w+ f^{\prime}_{1}\mu_{\eta}w^{\prime} \right),
\nonumber \\
m^{2}_{\eta^{\prime}_{1}\rho_{1}}&=&\frac{1}{6}\left(
f^{\prime}_{1}\mu_{\rho}w^{\prime}+f_{1}\mu_{\eta}w \right) ,\nonumber \\
m^{2}_{\eta^{\prime}_{1}\rho^{\prime}_{1}}&=&
\frac{f^{\prime}_{1}}{6}\mu_{\chi}w-k^{\prime}_{1}w^{\prime}.
\label{matrixmasselementstoplot}
\end{eqnarray}
We want to remind that the parameters $\mu_{\eta},
\mu_{\rho},\mu_{\chi},k_{1}$ and $k^{\prime}_{1}$ have mass
dimension, while $f_{1}$ and $f^{\prime}_{1}$ are dimensionless,
see Eq.(\ref{sp3susy2}).

Solving the characteristic equation, we have four massive fields
with the physical eigenvalues \bea m^{2}_{H^{0}_{1,2}} & = &
\pm\frac{1}{\sqrt{2}}\sqrt{\left(m_{\eta_1 \rho_1}^4+m_{\eta_1
\rho'_1}^4+m^4_{\eta'_1 \rho_1}+m_{\eta'_1 \rho'_1}^4 -\sqrt{M}
\right)},\crn m^{2}_{H^{0}_{3,4}} & = &
\pm\frac{1}{\sqrt{2}}\sqrt{\left(m_{\eta_1 \rho_1}^4+m_{\eta_1
\rho'_1}^4+m^4_{\eta'_1 \rho_1}+m_{\eta'_1 \rho'_1}^4 +\sqrt{M}
\right)}, \label{dm2}\eea where \be M=\left(m_{\eta_1
\rho_1}^4+m_{\eta_1 \rho'_1}^4+m^4_{\eta'_1 \rho_1}+m_{\eta'_1
\rho'_1}^4\right)^2
-4\left(m_{\eta_1\rho'_1}^2m_{\eta'_1\rho_1}^2-
m_{\eta_1\rho_1}^2m_{\eta'_1\rho'_1}^2\right)^2.
\label{sqrtdm2}
\ee
\subsection {Spectrum in neutral pseudoscalar sector}
 In the pseudoscalar sector,  after imposing the constraint
 equation, we have two Goldstone bosons$\chi_4, \chi'_4$. Those other
 have mixing square matrix in
 the base of $(\eta_2$, $\rho_2$, $\eta_2^\prime$,  $\rho_2^\prime$, $\eta_4$,
 $\eta^\prime_4$, $\chi_2$, $\chi^\prime_2)$,
 as follows
 \be
 M_{PH}^2 =\left (\ba{cccc}
  M_{4\eta'\rho'}^2
   & 0 & 0&0
 \\ 0 & m^2_{\eta_4}& 0 & 0  \\
 0 & 0 & m^2_{\eta^\prime_4}&0\\
 0 & 0 & 0 & M_{2\chi_2\chi_2^\prime}^2  \ea \right)
\ee
 with
\be M_{4\eta'\rho'}^2=\left (\ba{cccc} 0 & m^2_{\eta_2\rho_2} & 0&
m^2_{\eta_2\rho_2^\prime}
\\  m^2_{\eta_2\rho2} & 0 & m^2_{\eta^\prime_2\rho_2}&0 \\ 0&
m^2_{\eta^\prime_2\rho_2}  & 0 & m^2_{\eta^\prime_2\rho^\prime_2}
 \\ m^2_{\eta_2\rho^\prime_2} & 0 &  m^2_{\eta^\prime_2\rho^\prime_2} & 0\ea
 \right),
\label{mass1}\ee

\begin{equation}
M_{2\chi_2\chi^\prime_2}^2=\frac{g^2}{2}\left (\ba{cc} w^{\prime
2} &w w^{\prime}
\\ w w^{\prime}&w^{2} \ea
\right)\label{dm3}\end{equation}
 We also see that, the bileptons
  $ \eta_4, \eta^{\prime}_4 $ do not mix with others,
 and they are   physical fields with masses given by
 \bea
 m^2_{\eta_4} & = &  m^2_{\eta_3}, \label{dma3}\\
  m^2_{\eta^\prime_4}& = &  m^2_{\eta^\prime_3}.
 \label{dmas4}
 \eea
\hs The square mass matrix  $ M_{\chi_2 \chi'_2} $ satisfies
condition (\ref{dm1}). Thus, it gives us one Goldstone boson and
one physical massive field $m^2_{\zeta_{\chi_2 \chi_2^\prime}}$
with mass:
 \be
m^2_{\zeta_{\chi_2 \chi_2^\prime}}=m^2_{\zeta_{\chi_1
\chi_1^\prime}}.
 \ee
 From $4 \times 4$ mass matrix $M_{4\eta'\rho'}^2$, we get the following
matrix elements of Eq.(\ref{mass1})
\begin{eqnarray}
m^{2}_{\eta_{2}\rho_{2}}&=&-\frac{f_{1}}{6}\mu_{\chi}w^{\prime}+k_{1}w
= - m^{2}_{\eta_{1}\rho_{1}},\nonumber \\
m^{2}_{\eta_{2}\rho_{2}^{\prime}}&=&\frac{1}{6}\left(
f_{1}\mu_{\rho}w+ f^{\prime}_{1}\mu_{\eta}w^{\prime} \right)
=m^{2}_{\eta_{1}\rho_{1}^{\prime}}, \nonumber \\
m^{2}_{\eta^{\prime}_{2}\rho_{2}}&=&\frac{1}{6}\left(
f^{\prime}_{1}\mu_{\rho}w^{\prime}+f_{1}\mu_{\eta}w \right)=
m^{2}_{\eta^{\prime}_{1}\rho_{1}}, \nonumber \\
m^{2}_{\eta^{\prime}_{2}\rho^{\prime}_{2}}&=&
-\frac{f^{\prime}_{1}}{6}\mu_{\chi}w+k^{\prime}_{1}w^{\prime}=-
m^{2}_{\eta^{\prime}_{1}\rho^{\prime}_{1}}.
\end{eqnarray}
four massive pseudoscalar bosons with the same  mass as in the
scalar sector (\ref{dm2}).

To conclude this section, we note that,  the scalar sector
contains Goldstone bosons for $Z, Z'$ and $X^0$, $X^{0*}$.

\subsection{Spectrum in the charged scalar sector}
 In the charged sector, the $8\times8$  mass matrix in the basis
 of
 $\eta^{-},\rho^{-}_{1},\eta^{'-}, \rho^{-'}_{1},\rho^{-}_{2},\rho^{'-}_{2},
\chi^{-},\chi^{'-}$  has form as follows:
 \be
 M^{2}_{charge } =\left (\ba{cccc}
  M^{ 2 }_{4c\eta'\rho'}
   & 0 & 0&0
 \\ 0 & m^2_{\rho^-_2}& 0 & 0  \\
 0 & 0 & m^2_{\rho^{\prime-}_2}&0\\
 0 & 0 & 0 & M_{2\chi^-\chi^{\prime+}}^2  \ea \right)
\ee
 with
\be M_{4c\eta'\rho'}^2=\left (\ba{cccc} 0 & m^2_{\eta^-\rho^+_1} &
0& m^2_{\eta^-\rho_1^{\prime+}}
\\  m^2_{\eta^+\rho_1^-} & 0 & m^2_{\eta^{'+}\rho_1^-}&0 \\ 0&
m^2_{\eta^{\prime-}\rho^+_1}  & 0 &
m^2_{\eta^{\prime-}\rho^{\prime+}_1}
 \\ m^2_{\eta^+\rho^{\prime-}_1} & 0 &  m^2_{\eta^{\prime+}\rho^{\prime-}_1} & 0\ea
 \right),
\label{mass10}\ee

\begin{eqnarray}
m^2_{\eta^-\rho^+_1}&=&wk_1-\frac{1}{6}w'\mu_{\chi}f_1 =-
 m^2_{\eta_1 \rho_1},\\
m^2_{\eta^-\rho_1^{\prime+}}&=&- \frac{1}{6}\left (w
\mu_{\rho}f_1+w'\mu_{\eta}
f'_1\right)=- m^2_{\eta_1 \rho'_1}, \\
m^2_{\eta^+\rho_1^-}&=& k_1w-\frac{1}{6}w'\mu_{\chi}f_1
= m^2_{\eta^- \rho_1^+},  \\
m^2_{\eta^{'+}\rho_1^-}&=&-\frac{1}{6}\left ( w\mu_{\eta}f_1
+w'\mu_{\rho}f'_1
\right)= - m^2_{\eta'_1 \rho_1},\\
m^2_{\eta^{\prime-}\rho^+_1}&=&-\frac{1}{6}\left (
w\mu_{\eta}f_1+w'\mu_{\rho}
f'_1\right)= - m^2_{\eta'_1 \rho_1}, \\
m^2_{\eta^{\prime-}\rho^{\prime+}_1}&=&-\frac{1}{6}w
\mu_{\chi}f'_1+w'k_{1}^{\prime}= - m^2_{\eta'_1 \rho'_1},\\
m^2_{\eta^+\rho^{\prime-}_1}&=&-\frac{1}{6}\left (w
\mu_{\rho}f_1+w'\mu_{\eta}f'_1\right)= - m^2_{\eta_1 \rho'_1}, \\
m^2_{\eta^{\prime+}\rho^{\prime-}_1}&=&-\frac{1}{6}w
\mu_{\chi}f'_1+w'k_{1}^{\prime} = - m^2_{\eta'_1 \rho'_1},
\label{masschar11}
\end{eqnarray}
$M_{4c\eta'\rho'}^2$ gives
 four massive charged scalars having mass as (\ref{dm2}).

We have other couple
  $ \rho^+_2, \rho^{\prime-}_2 $ do not mix with others,
 they are   physical fields with masses given by
 \bea
 m^2_{\rho^-_2} & = & \frac{g^2}{2}\left(w^2-w^{'2}\right)-\frac{f_1^2w^2}{9},
\label{dma3}\\
  m^2_{\rho^{'-}_2}& = &
  -\frac{g^2}{2}\left(w^2-w^{'2}\right)-\frac{f^{'2}_1w^{'2}}{9}
  =  m^2_{\eta'_3}
 \label{dmas4}
 \eea

 The $M_{2\chi^-\chi^{\prime+}}^2$
\begin{equation}
M_{2\chi^-\chi^{\prime+}}^2=\frac{g^2}{2}\left (\ba{cc} w^{\prime
2} & -w w^{\prime}
\\ -w w^{\prime}&w^{2} \ea
\right),\label{dm4}\end{equation}
 also gives us one Goldstone
 bosons and one mass as (\ref{ln1}).

\section{Numerical analysis}
\label{sec:numerical}

We will use below the following set of parameters in the scalar
potential \cite{s331r}:
\begin{equation}
f_1=2,\quad f^\prime_1=10^{-3},\quad{\rm (dimensionless)}
\label{fs}
\end{equation}
and
\begin{eqnarray}
k_1&=&k^\prime_1=10, \quad \mbox{} [GeV],\\
\mu_\eta&=&\mu_\rho=10, \quad \mbox{} [GeV],\\
\mu_\chi&=&100, \quad \mbox{} [GeV]. \label{ks}
\end{eqnarray}
Here we  assume  that $v=1000$ [GeV], $v^\prime=1500$ [GeV].

Diagonalizing the matrices we got the mass eigenstates, and below we
present our results.

\subsection{Neutral real scalar}

The eigenvalues in (GeV) of $M_{4\eta\rho}^2$ given at Eq.(\ref{dm2})
\begin{eqnarray}
\sqrt{|m^{2}_{H^{0}_{1,2}}|}&=&123.2, \\
\sqrt{|m^{2}_{H^{0}_{3,4}}|}&=&200.5
\end{eqnarray}
Note that the values presented above are in agreement with the
current $95\%$ CL mass bound on the lightest scalar at MSSM which
is $91$ GeV \cite{pdg}.

To the case of $\eta_3, \eta'_3$ at Eqs.(\ref{dma1},\ref{dmas2})
the eigenvalues in (GeV) are
\begin{eqnarray}
\sqrt{|m^{2}_{H^{0}_{5}}|}&=&480.7, \label{e3ep3e4ep4v1} \\
\sqrt{m^{2}_{H^{0}_{6}}}&=&516.6, \label{e3ep3e4ep4v2}
\end{eqnarray}
respectively. On the case of the eigenvalues of
$M_{2\chi_1\chi^\prime_1}^2$ and $M_{2\chi_3\chi^\prime_3}^2$ at
Eq.(\ref{ma3}), we got in (GeV) the following value:
\begin{eqnarray}
\sqrt{m^{2}_{H^{0}_{7}}}&=&832.7 \label{2c1cp12c2cp2}
\end{eqnarray}
while to $M_{2\chi_3\chi^\prime_3}^2$ at Eq.(\ref{ma4}), we got
\begin{eqnarray}
\sqrt{m^{2}_{H^{0}_{8}}}&=&968.
\end{eqnarray}

\subsection{Neutral imaginary scalar}

To the eigenvalues of the matrix $M_{4\eta'\rho'}^2$, given at
Eq.(\ref{mass1}), we got the following values in (GeV)
\begin{eqnarray}
\sqrt{|m^{2}_{A^{0}_{1,2}}|}&=&120.6, \\
\sqrt{|m^{2}_{A^{0}_{3,4}}|}&=&201.1.
\end{eqnarray}
Our values again are in agreement with the mass bound on the
lightest pseudoscalar at MSSM  which is $91.9$ GeV \cite{pdg}.

On the other hand the eigenvalue of the massive pseudoscalar of
$M_{2\chi_2\chi_2^\prime}^2$ is given ate Eq.(\ref{2c1cp12c2cp2})
while to the case of $\eta_4, \eta'_4$ is given at
Eqs.(\ref{e3ep3e4ep4v1},\ref{e3ep3e4ep4v2}).

\subsection{Charged sector}

 To the eigenvalues of the matrix $M^{ 2 }_{4c\eta'\rho'}$, given at
Eq.(\ref{masschar11}), we got the following values in (GeV)
\begin{eqnarray}
\sqrt{|m^{2}_{H^{+}_{1,2}}|}&=&102, \\
\sqrt{|m^{2}_{H^{+}_{3,4}}|}&=&124.2
\end{eqnarray}
These values presented above are in accordance with the mass bound
on the lightest charged scalar at MSSM  which is $79.3$ GeV
\cite{pdg}.

To the case of $\rho^{\prime-}_2, \rho^{\-}_2$ at Eqs.
(\ref{dma3},\ref{dmas4})the
eigenvalues in (GeV) are
\begin{eqnarray}
\sqrt{m^{2}_{H^{+}_{5}}}&=&516.4, \\
\sqrt{|m^{2}_{H^{+}_{6}}|}&=&843.3, \label{r2rp2}
\end{eqnarray}
respectively. On the other hand the eigenvalue of the massive
pseudoscalar of $M_{2\chi_2\chi_2^\prime}^2$ is given at
Eq.(\ref{2c1cp12c2cp2}).

\section{Plots}
\label{sec:plots}

Using the values giving at Eqs.(\ref{fs}$\div$\ref{ks}), we are
going only changing $w$ and $w^{\prime}$ in our analysis. We get
the following mass elements matrix to
Eq.(\ref{matrixmasselementstoplot})

\begin{figure}[htbp]
\begin{center}
\vglue -0.009cm
\mbox{\psfig{file=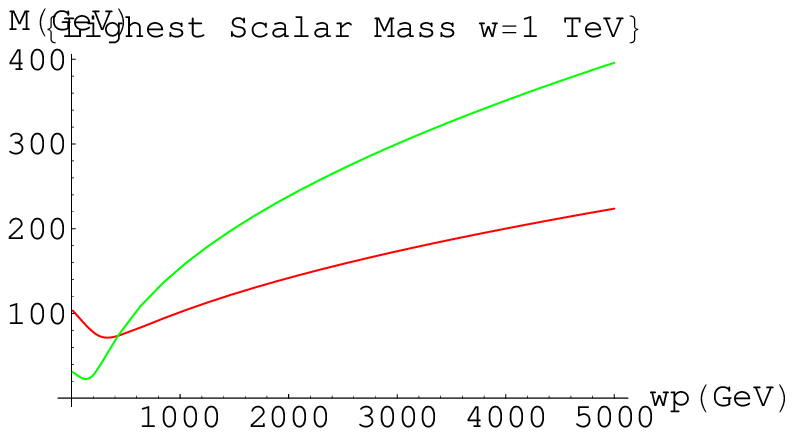,width=0.7\textwidth,angle=0}}
\end{center}
\caption{The masses  $\sqrt{m^{2}_{H^{0}_{1}}}$ (red lines), and
$\sqrt{m^{2}_{H^{0}_{3}}}$ (green lines) of the $4\times4$  matrix in
the neutral scalar as function of $w^{\prime}$.} \label{fig1a}
\end{figure}

\begin{figure}[htbp]
\begin{center}
\vglue -0.009cm
\mbox{\psfig{file=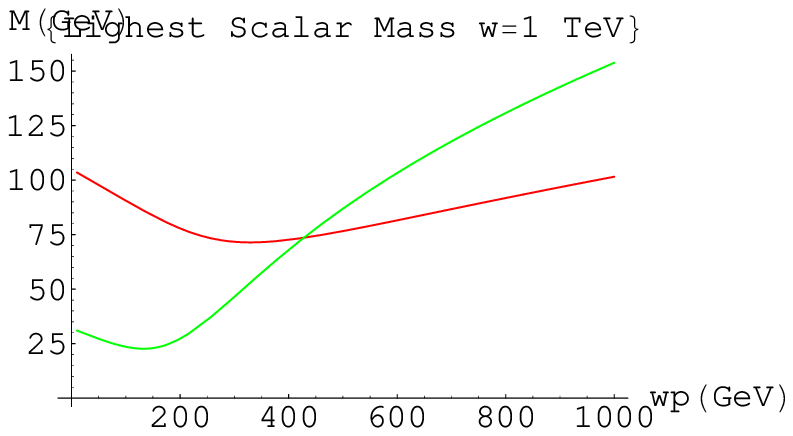,width=0.7\textwidth,angle=0}}
\end{center}
\caption{The masses  $\sqrt{m^{2}_{H^{0}_{1}}}$ (red lines), and
$\sqrt{m^{2}_{H^{0}_{3}}}$ (green lines) of the $4\times4$  matrix in
the charged scalar as function of $w^{\prime}$.} \label{fig1b}
\end{figure}

\begin{figure}[htbp]
\begin{center}
\vglue -0.009cm
\mbox{\psfig{file=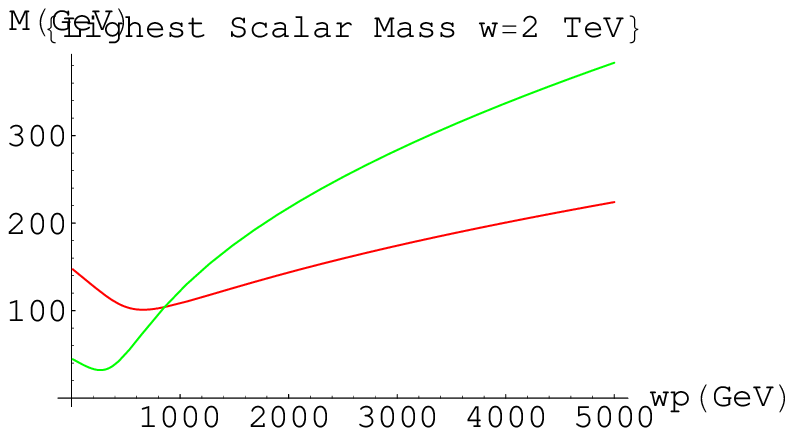,width=0.7\textwidth,angle=0}}
\end{center}
\caption{The masses  $\sqrt{m^{2}_{H^{0}_{1}}}$ (red lines), and
$\sqrt{m^{2}_{H^{0}_{3}}}$ (green lines) of the $4\times4$  matrix in
the neutral scalar as function of $w^{\prime}$.} \label{fig2}
\end{figure}

\begin{figure}[htbp]
\begin{center}
\vglue -0.009cm
\mbox{\psfig{file=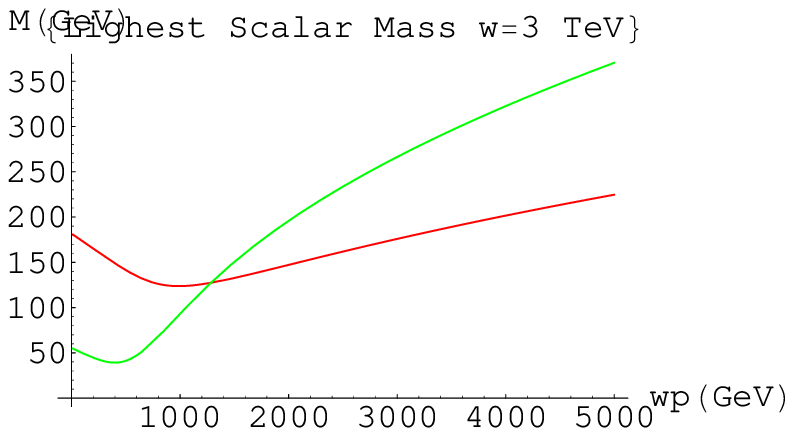,width=0.7\textwidth,angle=0}}
\end{center}
\caption{The masses  $\sqrt{m^{2}_{H^{0}_{1}}}$ (red lines), and
$\sqrt{m^{2}_{H^{0}_{3}}}$ (green lines) of the $4\times4$  matrix in
the neutral scalar as function of $w^{\prime}$.} \label{fig3}
\end{figure}

From the Fig.\ref{fig1a} the eigenvalue $\sqrt{m^{2}_{H^{0}_{1}}}$
at $w^{\prime}=5000$ GeV is $223.7$ GeV (Fig. \ref{fig1b} for $w'$
= 1 TeV), while in Fig.\ref{fig2} the same eigenvalue rise to
$224$ GeV, at the end, considering   Fig.\ref{fig3} we get $224.7$
GeV. Then, we can conclude that the lightest scalar mass of our
model has the upper limit around $230$ GeV.

\begin{figure}[htbp]
\begin{center}
\vglue -0.009cm
\mbox{\psfig{file=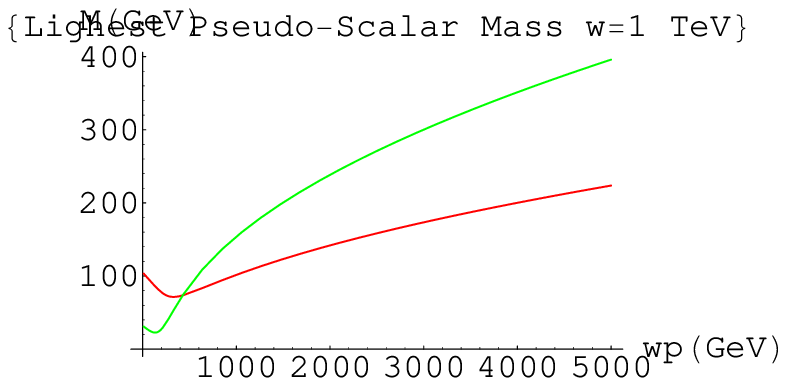,width=0.7\textwidth,angle=0}}
\end{center}
\caption{The masses  $\sqrt{m^{2}_{A^{0}_{1}}}$ (red lines), and
$\sqrt{m^{2}_{A^{0}_{3}}}$ (green lines) of the $4\times4$  matrix in
the neutral pseudo scalar as function of $w^{\prime}$.}
\label{fig4a}
\end{figure}

\begin{figure}[htbp]
\begin{center}
\vglue -0.009cm
\mbox{\psfig{file=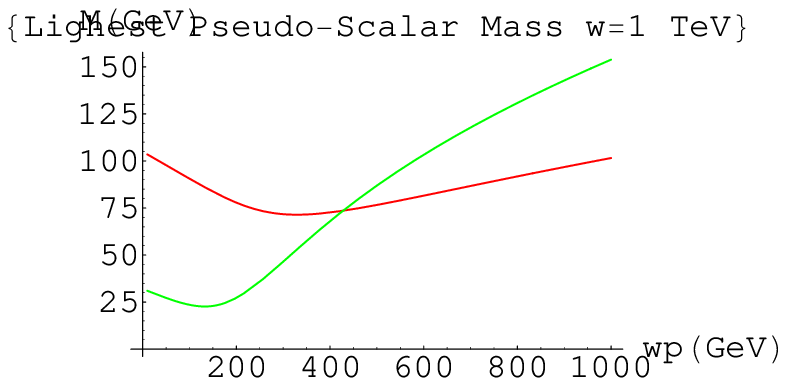,width=0.7\textwidth,angle=0}}
\end{center}
\caption{The masses  $\sqrt{m^{2}_{A^{0}_{1}}}$ (red lines), and
$\sqrt{m^{2}_{A^{0}_{3}}}$ (green lines) of the $4\times4$  matrix in
the charged scalar as function of $w^{\prime}$.} \label{fig4b}
\end{figure}

\begin{figure}[htbp]
\begin{center}
\vglue -0.009cm
\mbox{\psfig{file=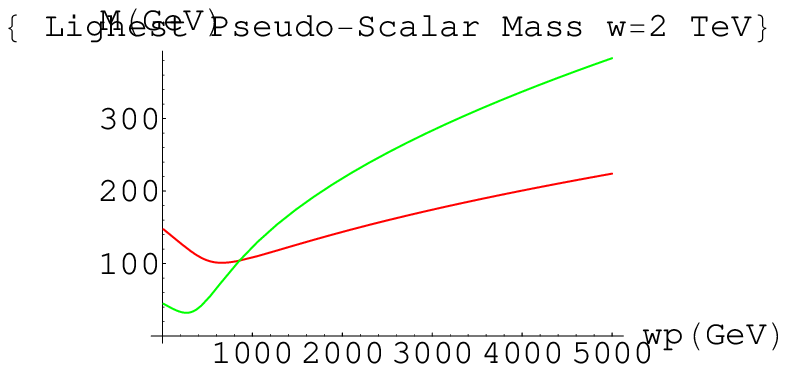,width=0.7\textwidth,angle=0}}
\end{center}
\caption{The masses  $\sqrt{m^{2}_{A^{0}_{1}}}$ (red lines), and
$\sqrt{m^{2}_{A^{0}_{3}}}$ (green lines) of the $4\times4$  matrix in
the neutral pseudo scalar as function of $w^{\prime}$.}
\label{fig5}
\end{figure}

\begin{figure}[htbp]
\begin{center}
\vglue -0.009cm
\mbox{\psfig{file=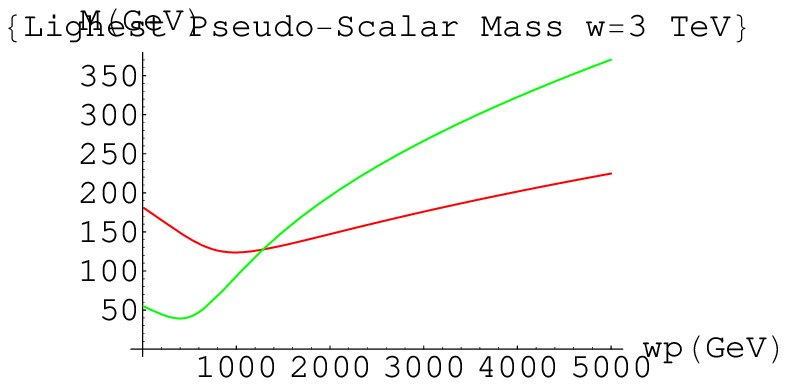,width=0.7\textwidth,angle=0}}
\end{center}
\caption{The masses  $\sqrt{m^{2}_{A^{0}_{1}}}$ (red lines), and
$\sqrt{m^{2}_{A^{0}_{3}}}$ (green lines) of the $4\times4$  matrix in
the neutral pseudo scalar as function of $w^{\prime}$.}
\label{fig6}
\end{figure}

From the Fig.\ref{fig4a} the eigenvalue $\sqrt{m^{2}_{A^{0}_{1}}}$
at $w^{\prime}=5000$ GeV is $223.7$ GeV (Fig. \ref{fig4b} for $w'$
= 1 TeV), while in Fig.\ref{fig5} the same eigenvalue rise to
$222$ GeV, at the end, considering Fig.\ref{fig6} we get $224.7$
GeV. Then, we can conclude that the lightest scalar mass of our
model has the upper limit around $230$ GeV.

\begin{figure}[htbp]
\begin{center}
\vglue -0.009cm
\mbox{\psfig{file=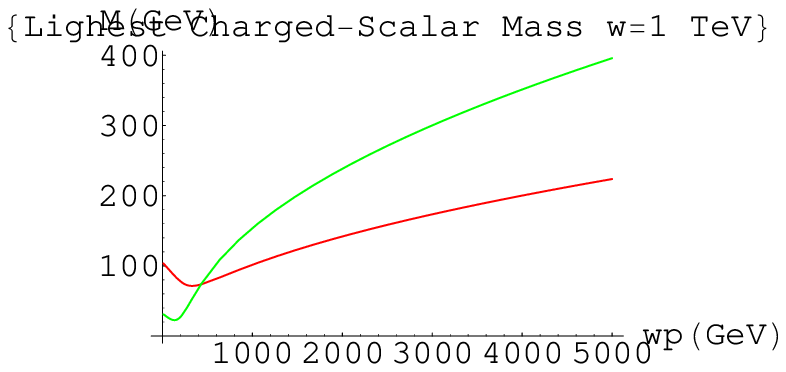,width=0.7\textwidth,angle=0}}
\end{center}
\caption{The masses  $\sqrt{m^{2}_{H^{+}_{1}}}$ (red lines), and
$\sqrt{m^{2}_{H^{+}_{3}}}$ (green lines) of the $4\times4$  matrix in
the charged scalar as function of $w^{\prime}$.} \label{fig7a}
\end{figure}

\begin{figure}[htbp]
\begin{center}
\vglue -0.009cm
\mbox{\psfig{file=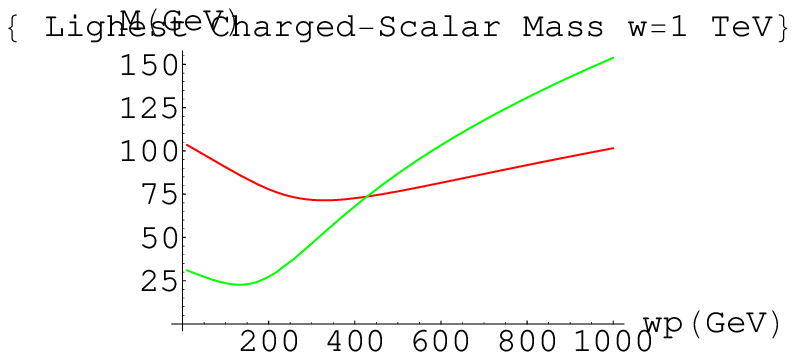,width=0.7\textwidth,angle=0}}
\end{center}
\caption{The masses  $\sqrt{m^{2}_{H^{+}_{1}}}$ (red lines), and
$\sqrt{m^{2}_{H^{+}_{3}}}$ (green lines) of the $4\times4$  matrix in
the charged scalar as function of $w^{\prime}$.} \label{fig7b}
\end{figure}

\begin{figure}[htbp]
\begin{center}
\vglue -0.009cm
\mbox{\psfig{file=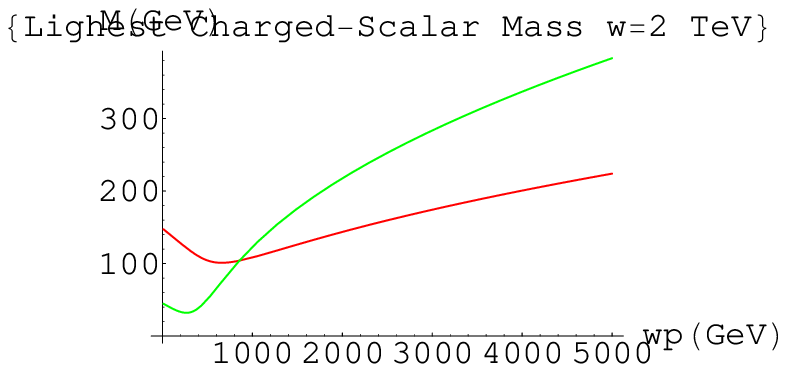,width=0.7\textwidth,angle=0}}
\end{center}
\caption{The masses  $\sqrt{m^{2}_{H^{+}_{1}}}$ (red lines), and
$\sqrt{m^{2}_{H^{+}_{3}}}$ (green lines) of the $4\times4$  matrix in
the charged scalar as function of $w^{\prime}$.} \label{fig8}
\end{figure}

\begin{figure}[htbp]
\begin{center}
\vglue -0.009cm
\mbox{\psfig{file=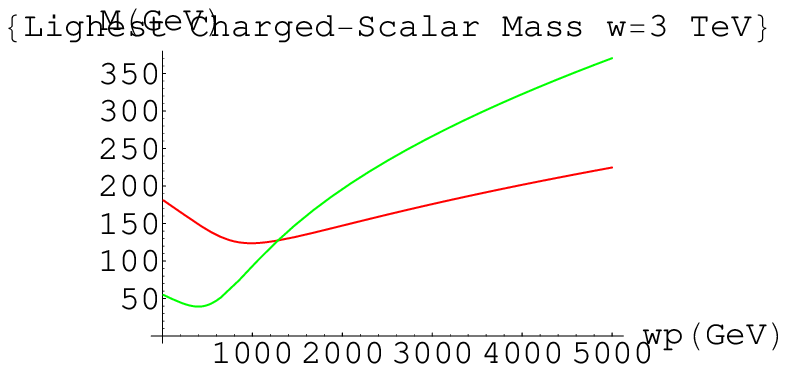,width=0.7\textwidth,angle=0}}
\end{center}
\caption{The masses  $\sqrt{m^{2}_{H^{+}_{1}}}$ (red lines), and
$\sqrt{m^{2}_{H^{+}_{3}}}$ (green lines) of the $4\times4$  matrix in
the charged scalar as function of $w^{\prime}$.} \label{fig9}
\end{figure}

From the Fig.\ref{fig7a} the eigenvalue $\sqrt{m^{2}_{H^{+}_{1}}}$
at $w^{\prime}=5000$ GeV is $223.7$ GeV (Fig. \ref{fig7b} for $w'$
= 1 TeV), while in Fig.\ref{fig8} the same eigenvalue rise to
$224$ GeV, at the end, considering Fig.\ref{fig9} we get $224.7$
GeV. Then, we can conclude that the lightest scalar mass of our
model has the upper limit around $230$ GeV.

From all our plots we can obviously see that the scalar mass are given
by the Eq.(\ref{dm2}), as we have mention
during this article.

\section{Conclusions}
\label{sec:conclusion}

On this article we constructed all the spectrum from the scalar
sector of the supersymmetric 3-3-1 model with RH neutrinos. We
show that there is no mixing between scalars having $L=0$ and
bilepton scalars having $L=2$. On this model we have six Goldstone
bosons: two in neutral sector, three in pseudo-scalar sector and
one in charged scalar sector. We analyze also, numerically, the
values of the masses of their physical mass spectrum of the
scalars. All the scalar sector of our model contain the upper
limit of $230$ GeV to the mass of the lightest scalar. All these
values are in agreement with the lower limit of the SM  Higgs
boson obtained by LEP.

The scalar sector of the non-supersymmetric 3-3-1 model was
studied at \cite{l97}, while the production of the standard model
Higgs boson at $pp$ colliders  was studied in Ref.\cite{longlinh}.
On this article, if the mass of $Z^{\prime}$ is not higher than 1
TeV then the process $pp \rightarrow hZ$ is observable at LHC in
the case $M_{h}<780$ GeV. If $M_{Z^{\prime}}$ is from 2 until 4
TeV then the process is observable, for $M_{h}<600$ GeV. This
analyze is still hold if we assume that the sparticles are heavier
than the usual ones.

At last the gauge boson production was analyzed \cite{zrh}, and on
this article a complete set of quadratic gauge boson coupling in
both 3-3-1 models was presented. The authors deduced that at tree
level the quartic divergences are canceled and then unitarity is
satisfied. This analyze is still hold on this model.

{\it Acknowledgement}:
 This work was supported in part
by National Council for
Natural Sciences of Vietnam contract No: KT  - 41064.\\[0.3cm]\\[0.3cm]

\newpage

\begin{thebibliography}{9}
\bibitem{singer}M. Singer, J. W. F. Valle, and J. Schechter, Phys.
Rev. D {\bf22}, (1980) 738
\bibitem{ppf} F. Pisano and V. Pleitez, Phys. Rev.  D {\bf46}, (1992) 410;
P. H. Frampton, Phys. Rev. Lett. {\bf69}, (1992) 2889; R. Foot
{\it et al,} Phys. Rev. D {\bf47}, (1993) 4158.
\bibitem{331rh} R. Foot, H. N. Long, and Tuan A.
Tran,
 Phys. Rev. D {\bf50}, (1994) R34; J. C. Montero, F. Pisano, and
 V. Pleitez, Phys. Rev. D {\bf 47}, (1993)
 2918; H. N. Long, Phys. Rev. D  {\bf 53}, (1996) 437;
Phys. Rev. D  {\bf 54}, (1996) 4691.
\bibitem{dongl2}  F. Pisano, Mod. Phys. Lett
A {\bf 11} 2639 (1996); A. Doff and F. Pisano, Mod. Phys. Lett A
{\bf 14} 1133 (1999); C. A. de S. Pires and O. P. Ravinez, Phys.
Rev. D {\bf 58} 035008 (1998); C. A. de S. Pires, Phys. Rev. D
{\bf 60} 075013 (1999);  P. V. Dong and H. N. Long, [arXiv:
hep-ph/0507155].
\bibitem{longvan} H. N. Long and V. T. Van, J. Phys. {\bf G25}, 2319 (1999).
\bibitem{longlan} D. Fregolente and M. D. Tonasse, Phys. Lett. B
{\bf 555}, 7 (2003);
 H. N. Long and N. Q. Lan, Europhys. Lett. {\bf 64}, 571
(2003).
\bibitem{s331r} J. C. Montero, V. Pleitez and M. C. Rodriguez,
Phys. Rev. D 70 (2004) 075004.

\bibitem{dias} A. G. Dias, C. A. de S. Pires, and P. S. Rodrigues
da Silva, Phys. Rev. D68, 115009 (2003).

\bibitem{haber} H. E. Haber and G. L. Kane, Phys. Rep. {\bf 117}, 75 (1985).
\bibitem{wb} J. Wess and J. Bagger, {\it Supersymmetry and
Supergravity}, 2nd edition, Princeton University Press, Princeton
NJ, (1992).
\bibitem{10}L. Girardello and  M. T. Grisaru, {\em Nucl.\  Phys.\ } {\bf B194}
(1982) 65.
\bibitem{l97} H. N. Long, Mod. Phys. Lett. A 13 (1998) 1865.
\bibitem{longinami} H. N. Long and T. Inami, Phys. Rev. {\bf D 61},
 075002 (2000).
\bibitem{pdg} S. Eidelman {\it et al.}, Phys. Lett. B {\bf 592} (2004).
\bibitem{longlinh}Le Duc Ninh and H. N. Long, [arXiv:hep-ph/0507069].
\bibitem{zrh}D. T. Binh, D. T. Huong, T. T. Huong, H. N. Long and D.V. Soa,
J. Phys. G {\bf 29}, 1213 (2003).
\end{thebibliography}
\end{document}